\documentclass[prl,twocolumn,showpacs,epsfig,longeq]{revtex4}
\usepackage{graphicx}
\usepackage{dcolumn}
\usepackage{amssymb}
\usepackage{bm}
\usepackage{epsfig}
\usepackage{psfrag}

\def\3dots{\:\raisebox{-0.5ex}{$\stackrel{\textstyle.}{:}$}\:}
\def\beq{\begin{equation}}
\def\eeq{\end{equation}}
\def\bea{\begin{eqnarray}}
\def\eea{\end{eqnarray}}

\begin{document}

\title{Symmetry properties of the large-deviation function of the velocity of a self-propelled polar particle}

\author{Nitin Kumar, Sriram Ramaswamy and A.K. Sood}
\affiliation{Department of Physics, Indian Institute of Science, Bangalore 560012, India}

\date{\today}
%\draft
\begin{abstract}
A geometrically polar granular rod confined in 2-D geometry, subjected to a sinusoidal vertical oscillation, undergoes noisy self-propulsion in a direction determined by its polarity. When surrounded by a medium of crystalline spherical beads, it displays substantial negative fluctuations in its velocity. We find that the large deviation function (LDF) for the normalized velocity is strongly non-Gaussian with a kink at zero velocity, and that the antisymmetric part of the LDF is linear, resembling the fluctuation relation known for entropy production, even when the velocity distribution is clearly non-Gaussian. We extract an analogue of the phase space contraction rate and find that it compares well with an independent estimate based on the persistence of forward and reverse velocities.
\end{abstract}
\maketitle

When a particle moves under the influence of a driving force through a noisy medium, it occasionally moves in the direction opposite to the force. The probability of such entropy consumption, relative to production, has been shown to obey the well-known fluctuation relations \cite{reviews}, which are a symmetry property of the large-deviation function (LDF) \cite{touchette} of the entropy production rate. Experiments on a surprising range of nonequilibrium systems find behaviour consistent with the FR \cite{carberry,wang, ciliberto1,science,nature,Goldburg,ciliberto2,Douarche,Shang,Ciliberto3,menon,puglisi,asood}, including athermal systems \cite{asood} where the noise is a consequence of the driving.. 

In this paper, we study the \textit{velocity} statistics of a geometrically polar particle in a dense monolayer of beads on a vertically agitated horizontal surface. The continuous input of energy through mechanical vibration is balanced by dissipation into the macroscopic number of internal degrees of freedom of each particle. Agitation feeds energy into the tilting vertical motion of the polar particle, which transduces it, via frictional contact with the base, into horizontal movement in a direction determined by its orientation in the plane \cite{sano,kudrolli,vjnmsr,sractivematterreview,dauchot}. The particle thus behaves like a noisy self-propelled object, with a statistically significant tendency to move in the ``reverse'' direction, i.e., opposite to its mean direction of  spontaneous motion. We are interested in large deviations of the velocity. Accordingly, let $P(W_{\tau})$ be the probability density of $W_{\tau}(t)=(1/{\tau})\int_t^{t+\tau}\!\left[V(t')/\left\langle V\right\rangle\right]dt'$ where $V(t) \equiv \mathbf{v}(t)\cdot\hat{\mathbf{n}}(t)$ with $\mathbf{v}(t)$ and $\hat{\mathbf{n}}(t)$ are the particle velocity and orientation vector in the plane, and $\langle \rangle$ denotes an average over the time $t$. The LDF is then $F(W_{\tau}) \equiv \lim_{\tau \to \infty}(-1/\tau)\ln P(W_{\tau})$. If $W_{\tau}$ were the entropy-production rate the FR would read $F(W_{\tau}) - F(-W_{\tau}) \propto W_{\tau}$. 

\begin{figure}
\centerline{\includegraphics[width=0.45\textwidth]{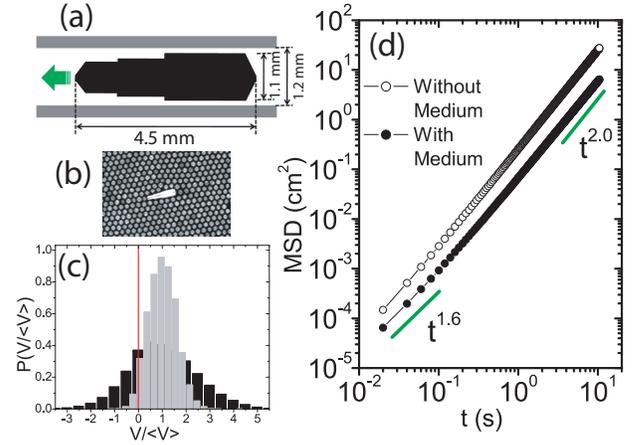}}
\caption{(a) Geometry of the polar particle: the fat arrow indicates the mean direction of ``self-propelled'' motion. (b) A typical experimental screen-shot showing the particle finding its way through a crystalline medium of 0.8 mm aluminium beads. (c) Comparison of normalized velocity distributions of polar particle without (gray bars) and with (black bars) medium. (d) Mean-square displacement (MSD) with and without medium particles. Short-time exponent (with medium) and long-time exponents (for both cases) displayed.}
\label{polar-particle}
\end{figure}

Our experimental results are as follows: (i) The particle velocity statistics satisfy a large-deviation principle. We are able to extract the LDF, $F(W_{\tau})$, and find that the corresponding probability is strongly non-Gaussian, with a kink at zero \cite{mehlfootnote}. (ii) The antisymmetric part $F(W_{\tau}) - F(-W_{\tau}) \propto W_{\tau}$, i.e., the velocity obeys the analogue of a fluctuation relation. (iii) From the velocity statistics, we calculate an analogue of phase-space contraction rate and show that it correlates very well with an independent estimate based on the difference in the persistence rates of negative and positive velocities. The need of such an independent estimate for contraction rates, in a fluctuation-relation context, has been emphasized recently \cite{cilibertostatphys}.

Two clarifications are essential here: (a) In principle, we are not measuring the LDF for the entropy production rate, as the distributions of power and velocity are distinct for general time-dependent driving; indeed, we have no access to the time-series of the propulsive force. We know of no earlier reports of symmetry relations analogous to the FRs for large deviations of the \textit{velocity} \cite{seifertframeinv}. (b) Such a relation would hold trivially if the velocity statistics happened to be Gaussian. We will see that the velocity statistics of our particle is far from Gaussian.  

Our experimental cell is a shallow circular geometry, of diameter $D =13$ cm, made of hardened aluminium alloy. To confine the particles to two dimensions, a glass lid is fixed on the external perimeter of the circle at a height of 1.2 mm above the base. We ensured that base and lid were uniformly flat to within 10 $\mu$m accuracy. This geometry is mounted on a permanent magnet shaker (LDS 406/8) and is shaken at a fixed frequency $f=200Hz$ and amplitude $a_{0}$ between 0.019 mm and 0.047 mm. The resulting accelerations $\Gamma \equiv {a_{0}(2 \pi f)^{2}}/{g}$, measured by an accelerometer (PCB Piezotronics 352B02) and nondimensionalized by gravity $g$, lie between 3.0 and 7.5. We ensured that our apparatus was level to the accuracy of a spirit-level; our results are insensitive to small deviations in levelling. The dynamics of the particles is recorded by a  high-speed digital camera (Redlake MotionPro X3) mounted vertically above the plate, with a maximum resolution of 1024$\times$1280. The frame rate used was 50 fps for all the results presented. The resolution of the resulting image was 0.1 mm which corresponds to a pixel length. The images were analyzed using ImageJ \cite{ImageJ}.

Our ``self-propelled'' polar particle is a brass rod, 4.5 mm long and 1.1 mm in diameter at its thick end as shown in Fig. \ref{polar-particle}(a). On a bare surface the dynamics is as follows: When confined between horizontal plates and shaken at $\Gamma > 4.5$ it moves on average 
% \cite{spudich} 
with narrow end forward along the arrow in Fig. \ref{polar-particle}(a), which we term the positive velocity direction. We determined the position and, thanks to the shape asymmetry of the particle, the orientation of the particle in each frame \cite{image_analysis}, and extracted the instantaneous particle velocity along its axis. Gray bars show a typical distribution of normalized velocity ($V/\langle V\rangle$) of this particle at $\Gamma$=7.5 in Fig. \ref{polar-particle}(c), clearly showing the tendency towards systematic directed motion with exceedingly rare backsteps. Below $\Gamma$=4.5 the same particle gradually starts showing a substantial number of negative velocity events whose significance we will discuss later.

\begin{figure}
\centerline{\includegraphics[width=0.45\textwidth]{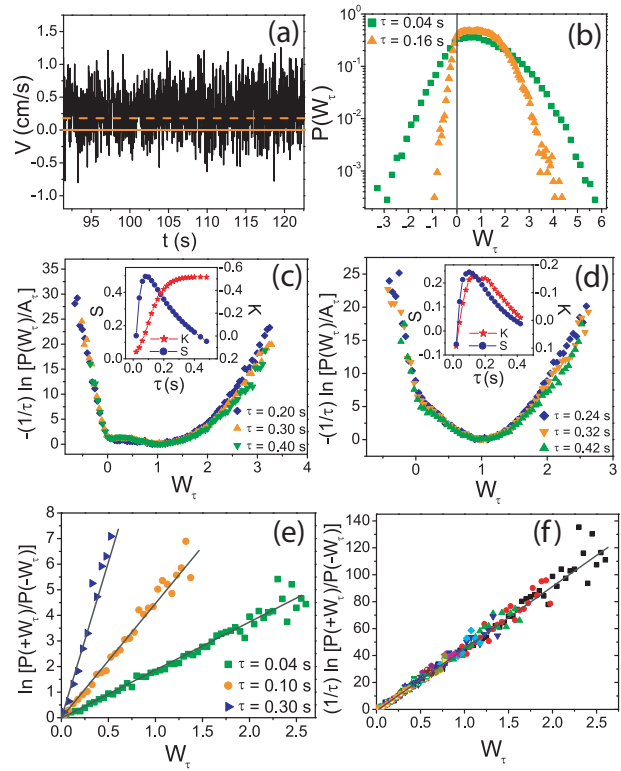}}
\caption{(a) Typical velocity fluctuations of polar particle in bead medium, with dashed line indicating mean velocity. All data except in (d) is for $\Gamma = 7.5$, $\Phi = 0.83$. (b) Probability distribution of $W_{\tau}$ for various $\tau$. Note exponential negative tail and flat peaks. (c) The collapsed large-deviation function. Inset shows kurtosis (K) and skewness (S) plotted as a function of $\tau$ clearly indicating the strongly non-Gaussian nature of the distributions. (d) Large-deviation function shown for $\Gamma = 6.5$ and $\Phi = 0.83$ and corresponding kurtosis (K) and skewness (S). (e) Linear dependence of $\ln {P(+W_{\tau})}/{P(-W_{\tau})}$ on $W_{\tau}$; $\tau$=0.04 s, 0.10 s and 0.30 s and (f) Data collapse of $({1}/{\tau}) \ln {P(+W_{\tau})}{P(-W_{\tau})}$ vs $W_{\tau}$ onto a single line for all $\tau \geq 0.04 s$ for $\Gamma = 7.5$ and $\Phi = 0.83$.}
\label{ldf-kink}
\end{figure} 

We then studied the motion of the polar particle when the experimental plate is filled with a close-packed monolayer of aluminium beads of diameter $d= 0.8$ mm. Note that the 1.2 mm gap thickness between base and lid leaves a clearance of 0.4 mm above the beads, and 0.1 mm above the thick end of the polar rod, allowing the play that keeps the system ``alive''. We study the velocity statistics of the polar particle as it pushes its way through a medium at area fraction $\Phi \equiv N (d/D)^2 \geq 0.8$ where $N$ is the total number of beads. At these concentrations the beads form a triangular lattice (Fig. \ref{polar-particle}(b)). Black bars in Fig. \ref{polar-particle}(c) show the normalized velocity distribution of the polar particle at $\Gamma = 7.5, \, \Phi$=0.83. Note the significant weight at negative velocities. Fig. \ref{polar-particle}(d) shows the mean square displacement (MSD) plot of the polar particle with and without the medium. We see that the particle which moved ballistically at all time scales in the absence of a medium shows, when surrounded by beads, a sub-ballistic short-time motion (MSD $\sim t^{1.6}$) which we attribute to short-time negative-velocity events.

Typical fluctuations in the velocity of the particle are shown in the time-series in Fig. \ref{ldf-kink}(a), which has an autocorrelation time $<0.02$ s. Approximately 50,000 frames were captured and the velocity obtained from every pair of successive frames. Negative events are clearly visible. After evaluating $W_{\tau}$ and corresponding probability $P(W_{\tau})$, our aim is to obtain an LDF $F(W_{\tau})$. To this end we construct the time series of $W_{\tau}$, dividing the $V(t)/\left\langle V(t)\right\rangle$ series into different bins of length $\tau$ and averaging over overlapping bins where the centre of each bin is shifted from the previous one by a time difference $0.02$ s to improve statistics. The results presented here are not sensitive to the value of the time-difference used in the analysis. Fig. \ref{ldf-kink}(b) shows the probability distribution for $W_{\tau}$ for $\tau$ = 0.04 s and 0.16 s. The distributions are highly non-Gaussian with exponential tails on the negative side.

To bring out the non-Gaussian nature, we calculate the skewness $S = \left\langle \delta W_\tau^3\right\rangle/\sigma ^3$ and kurtosis $K = \left\langle \delta W_\tau^4\right\rangle/\sigma^4 - 3$, where $\delta W_{\tau} = W_{\tau} - \langle W_{\tau} \rangle$ and $\sigma^2 = \langle \delta W_{\tau}^2\rangle$. Note that $S = 0 = K$ for a Gaussian distribution. It is clear from the inset to Fig. \ref{ldf-kink}(c) that the distributions of $W_\tau$ are highly skewed towards positive values at low $\tau$, and become flatter around the mean at large $\tau$.
\begin{figure}[!t]
\centerline{\includegraphics[width=0.45\textwidth]{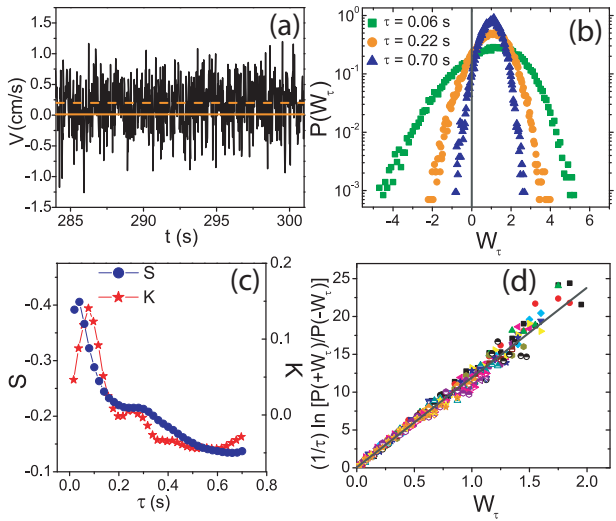}}
\caption{(a) Typical velocity time series of polar particle \textit{without} bead medium, at $\Gamma=3.0$; dashed line again indicates mean velocity. (b) Probability distribution of $W_{\tau}$ for $\tau$ = 0.06s, 0.22s and 0.70 s. (c) kurtosis ($K$) and skewness ($S$) as a function of $\tau$ indicating tendency to become more Gaussian at higher $\tau$. (d) $(1/\tau) \ln P(+W_{\tau}) / P(-W_{\tau})$ vs. $W_{\tau}$, shows a collapse only for $\tau\geq0.20 s$.} 
\label{no-med}
\end{figure}

In order to extract the LDF from our data, we begin by writing $P(W_{\tau})=A_{\tau}\exp{[-\tau F(W_{\tau})]}$, allowing for a prefactor $A_{\tau}$ independent of $W_{\tau}$ and varying more slowly than $\exp({-\tau})$. As $F$ is expected \cite{touchette} to vanish at the most probable or typical value of $W_{\tau}$, we estimate $A_{\tau}$ by the maximum value of $P(W_{\tau})$. We evaluate $\left(-1/\tau \right) \ln\left[P(W_{\tau})/A_{\tau}\right]$, find data collapse for $\tau>0.12 s$, and obtain the LDF. Fig. \ref{ldf-kink}(c) shows $F(W_{\tau})$ for $\tau$ = 0.20s, 0.30 s \& 0.40 s covering almost the entire range of $W_{\tau}$. $F(W_{\tau})$ shows a sharp kink at zero, remaining almost flat between 0 and 1. Fig. \ref{ldf-kink}(d) shows the LDF obtained for yet another case, $\Gamma = 6.5$ at $\Phi = 0.83$. The distribution is again non-Gaussian though $K$ and $S$ are low (see inset in Fig. \ref{ldf-kink}(d)), and a kink at zero can be seen. We have no explanation for the difference in the behaviours of $K$ in Figs. \ref{ldf-kink} (c) and (d).

A comparison is in order to the theoretically predicted LDF of the entropy production rate for a colloidal particle driven by a constant force through a periodic potential \cite{mehletal}. The resemblance of our LDF (Fig. \ref{ldf-kink}(c)) to their Fig. 1, bottom row, centre panel, is striking. Note however that we are measuring the LDF for the velocity, not the entropy production as in \cite{mehletal}. If the motion of our polar particle can be approximated as propelled by a constant force, then our results can be viewed as a confirmation of the predictions of \cite{mehletal}, albeit in a slightly more complicated medium. If it turns out that the propulsive force in our case has significant time-dependence, our results are all the more intriguing. At present however we have no independent way of obtaining a time-series for the force on the particle. 

We now examine the relative probabilities of positive and negative coarse-grained normalized velocities $W_{\tau}$. We find that $\ln \left[{P(+W_{\tau})}/{P(-W_{\tau})}\right]$ is linear in $W_{\tau}$, as shown in Fig. \ref{ldf-kink}(e) for $\tau$=0.04s, 0.10 and 0.30s  (only three $\tau$ values shown for clarity). This linearity persists to the highest $\tau$ values where the distribution is clearly non-Gaussian. Moreover, Fig. \ref{ldf-kink}(f) shows that $({1}/{\tau}) \ln \left[{P(+W_{\tau})}/{P(-W_{\tau})}\right]$ vs. $W_{\tau}$ collapses onto a single straight line for all $\tau$ values; the slope $\alpha$ = 43.7$\pm$5.2 $s^{-1}$. We find a similar result for other values of $\Gamma$ and $\Phi$. We conclude that the antisymmetric part of the LDF is linear in $W_{\tau}$; i.e $F(W_{\tau}) - F(-W_{\tau}) \propto W_{\tau}$. By analogy with the Gallavotti-Cohen SSFR \cite{reviews} for entropy flux, we are tempted to suggest a fluctuation relation for the particle's normalized velocity: $\lim_{\tau \to \infty} ({1}/{\tau})\ln {P(+W_{\tau})}/{P(-W_{\tau})}= \alpha W_{\tau}$. $\alpha$ may be regarded as similar to the phase space contraction rate in the conventional SSFR \cite{reviews}; we return to this point later.

\begin{figure}[!t]
%\hspace{+0.3cm}
\centerline{\includegraphics[width=0.45\textwidth]{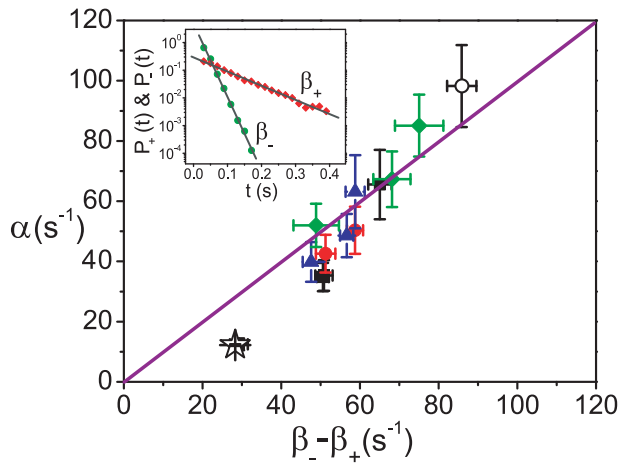}}
\caption{Analogue of phase-space contraction rate $\alpha$ vs. $\beta_- - \beta_+$, the difference in persistence rates of negative and positive velocities. Filled squares: $\Phi$=0.83 ($\Gamma$ = 7.5, 6.5, 5.5, 4.5); filled circles: $\Phi$=0.82 ($\Gamma$ = 7.5, 5.5); filled triangles: $\Phi$=0.81 ($\Gamma$ = 7.5, 6.5, 5.5); filled diamonds: $\Phi$=0.80 ($\Gamma$ = 7.5, 6.5, 5.5). The values for the case of a bare substrate at $\Gamma$ = 3.0 (open star) and the circular $1d$ channel at $\Gamma$ = 7.5 (open circle) are also shown. The solid line depicts $\alpha$ = ($\beta_- - \beta_+$). Inset: log-linear plot of $P_+(t)$ and $P_-(t)$ vs. time $t$, with slopes  $\beta_+$ and $ \beta_-$ respectively.}
\label{gammabeta}
\end{figure}

For comparison we repeat our experiment \textit{without} the bead medium, working at a lower  $\Gamma$=3.0 where the particle on a bare plate shows frequent negative velocity events \cite{low-gamma}. Fig. \ref{no-med}(a) shows the velocity time-series and Fig. \ref{no-med}(b) shows the distribution of $W_{\tau}$ for $\tau = 0.06$s, $0.22$s and $0.70$s. The trend seen here, contrary to that with the bead medium, is of a non-Gaussian distribution at lower $\tau$ = 0.06 s, becoming progressively more Gaussian as we integrate the time-series for higher $\tau$. It is clear from Fig. \ref{no-med}(c) that both skewness and kurtosis decrease with $\tau$. The quantity $(1/\tau W_{\tau}) \ln \left[{P_{\tau}(+W)}/{P_{\tau}(-W)}\right]$ approaches a $\tau$-independent constant only for $\tau\geq0.20 s$ (see Fig. \ref{no-med}(d)) where both non-Gaussian parameters are significantly low. This is unlike the case where the particle moves through the bead-bed, where this behavior persisted even when the distribution was noticeably  non-Gaussian.

Returning to the case with a bead-bed, we repeated the experiment for various combinations of $\Phi$ and $\Gamma$. We found that in all cases the antisymmetric part of the LDF for the velocity was linear, and obtained a range of $\alpha$ values. We now suggest that the $\alpha$ can be estimated independently, without reference to the LDF, as follows: We extract the probability $P_+(t)$ that a particle moving with positive velocity at time $0$ continues to do so upto time $t$, and similarly $P_-(t)$ for negative velocity. Each is found to decay exponentially, with rates $\beta_+$ and $\beta_-$ respectively as shown in Fig. \ref{gammabeta} inset. Since $\beta_-$ and $\beta_+$ respectively measure the mean rates of escape from regions of negative (atypical) and positive (typical) velocity, it seems plausible that the overall relaxation rate of the system should be the difference of the two, i.e. $\alpha = \beta_- - \beta_+$. Indeed, Fig. \ref{gammabeta} shows a convincing correlation between $\alpha$ and $\beta_-  - \beta_+$ for a range of $\Phi$ and $\Gamma$. The data for a particle moving at $\Gamma = 3.0$ on a bare plate, as well as that for motion at $\Gamma = 7.5$ in the one-dimensional circular track mentioned above, were also analyzed in the same manner (Fig. \ref{gammabeta}) and confirm the correlation between $\alpha$ and $\beta_-  - \beta_+$.

In conclusion, we have shown that a geometrically polar particle, when energized by vertical vibration and immersed in an array of spherical beads, displays frequent steps in the sense opposite to its mean direction of spontaneous motion. The resulting velocity distribution is highly non-Gaussian, and the large deviation function (LDF) shows a kink at zero velocity, as in \cite{mehletal}. Most intriguingly, the antisymmetric part of the LDF is linear, i.e., the velocity fluctuations obey a symmetry relation analogous to those known \cite{reviews} for the entropy production rate. We provide an independent estimate of the analogue of a phase-space contraction rate, in terms of persistence rates of positive and negative velocities. 

We thank O. Dauchot for valuable comments. AKS and SR respectively acknowledge support from a CSIR Bhatnagar fellowship and a DST J C Bose Fellowship.


\begin{thebibliography}{10}
\bibitem{mehlfootnote} We discuss later in the paper the relation of our findings to the study in \cite{mehletal}.
\bibitem{image_analysis}
We use ``erosion'' in ImageJ to calculate the orientation of the particle. Eroding an asymmetric object removes a layer of pixels from the surface of the object which causes a shift in centroid of the object towards the thick end. The vector joining the centroids of the original and the eroded object gives the orientation of the particle.
\bibitem{low-gamma}
We cannot perform this analysis at $\Gamma > 4.5$ as there are no negative events. Moreover, the velocity distribution at $\Gamma=7.5$ is too narrow to carry out an SSFR analysis (see Fig. \ref{polar-particle}(c) gray bars).
\end{thebibliography}
\end{document}